\documentclass[a4paper]{jpconf}
\usepackage{graphicx}
\usepackage[square,sort&compress,numbers]{natbib}
\usepackage[colorlinks=true,urlcolor=blue,anchorcolor=blue,citecolor=blue,filecolor=blue,linkcolor=blue,menucolor=blue,linktocpage=true,unicode=true,bookmarksopen=true,pdfa=true]{hyperref}
 
\usepackage{amsmath}
\usepackage{amssymb}
\usepackage{bbold}
\usepackage{slashed}
\usepackage{mathtools}
\usepackage{lmodern}
\usepackage[T1]{fontenc}
\usepackage{bbm}
\DeclareMathAlphabet{\mathbfi}{OML}{cmm}{b}{it}
\usepackage{color}
\usepackage{caption}
\usepackage{subcaption}

\let\originalleft\left
\let\originalright\right
\renewcommand{\left}{\mathopen{}\mathclose\bgroup\originalleft}
\renewcommand{\right}{\aftergroup\egroup\originalright}

\makeatletter

\def\@spliteq#1{\begin{equation}\begin{split}#1\end{split}\end{equation}}
\def\splitequation{\collect@body\@spliteq}

\makeatother

\renewcommand{\vec}[1]{{\ifnum9<1#1\mathbf{#1}\else\ifcat\noexpand#1\relax\boldsymbol{#1}\else\mathbfi{#1}\fi\fi}}

\newcommand{\mathi}{\mathrm{i}}
\let\oldre\Re
\let\oldim\Im
\renewcommand{\Re}{\oldre\mathfrak{e}\,}
\renewcommand{\Im}{\oldim\mathfrak{m}\,}

\newcommand{\measure}{\frac{{\rm d}^3p}{\freq}}
\newcommand{\freq}{\omega_p}

\makeatletter
\def\expect{\@ifnextchar[{\@expecttw@}{\@expect@ne}}
\def\@expecttw@[#1]#2{\left\langle{#2}\right\rangle^\text{#1}}
\def\@expect@ne#1{\left\langle{#1}\right\rangle}
\makeatother

\def\mailto#1{\href{mailto:#1}{\texttt{#1}}}

\usepackage{tikz}
\usepackage{tikz-feynman}

\begin{document}
\title{Formal developments in curved momentum space: the quantum field theory roadmap}

\author{S~A~Franchino-Vi{\~n}as\ad{1}}

\address{\ad{1} Helmholtz-Zentrum Dresden-Rossendorf, Bautzner Landstraße 400,\\\hspace{8pt}01328 Dresden, Germany.}

\vspace*{5pt}
\address{E-mail: \mailto{s.franchino-vinas@hzdr.de}}

\begin{abstract}
We advocate that the dual picture of spacetime noncommutativity , i.e. the existence of a curved momentum space, 
could be a way out to solve some of the open conceptual problems in the field, such as the basis dependence of observables. 
In this framework, we show how to build deformed Klein--Gordon and Dirac equations. 
In addition, we give an outlook of how one could define quantum field theories, both free and interacting ones.

\end{abstract}

\section{Introduction}
\label{sec:introduction}

After more than a century of independent successes, the theory of general relativity and that of the quantum are still divorced. 
Despite the efforts devoted to the various lines of research currently active in the field of quantum gravity, 
it is not clear how the latter framework will assist in mitigating the cosmological tensions plaguing our current understanding of the universe~\cite{Abdalla:2022yfr},
thus providing a fertile field to phenomenology~\cite{Addazi:2021xuf}.

A feature shared by most of the theories of quantum gravity is the appearance of a granular structure of spacetime in the high-energy limit, 
a fact that is gracefully encoded in noncommutative theories through a nonvanishing commutator of coordinates~\cite{Chamseddine:2022rnn, Douglas:2001ba}.
In this sense, noncommutative theories provide a promising playground for analyzing quantum gravity effects at the lowest possible energy scales.

From an algebraic perspective, great progress has been made since the seminal work of Snyder~\cite{Snyder:1946qz,Snyder:1947nq}.
Around thirty years ago,  the so-called $q$-deformed Poincaré algebra was introduced through a  Drinfeld--Jimbo deformation~\cite{Lukierski:1991pn},
which led an expansion in the study of the $\kappa$-Poincaré and Snyder algebras \cite{Kosinski:2001ii,Kowalski-Glikman:2003qjp, Mignemi:2008kn,Govindarajan:2009wt, Mignemi:2011gr,Poulain:2018two,Arzano:2020jro,Lizzi:2021rlb}. 

As is well known, a dual perspective to noncommutativity in spacetime if offered by curved momentum spaces~\cite{Kowalski-Glikman:2002oyi, AmelinoCamelia:2011pe, Carmona:2019fwf}. 
Indeed, if momentum operators commute, they offer the possibility of introducing a representation of the algebra in a Hilbert space built out of the eigenstates of the momenta operators. 
Numerous works have recently followed this path, studying for example the momentum space of $\kappa$-Minkowski~\cite{Lizzi:2020tci, Arzano:2010jw}, the relation of the momentum space with generalized uncertainty principle theories~\cite{Wagner:2021bqz} and Casimir energies~\cite{Franchino-Vinas:2020umq}.

From a conceptual point of view, the notion of a curved momentum space sheds light on some obscure points of noncommutative theories. 
First, depending on the symmetries of the momentum space, one could well accommodate doubly special relativity theories~\cite{Kowalski-Glikman:2002oyi, AmelinoCamelia:2011pe, Carmona:2019fwf}, i.e. theories in which the Lorentz group is deformed but not lost~\cite{AmelinoCamelia:2008qg}. 

Second, the principle of relative locality~\cite{Amelino-Camelia:2011hjg,Mercati:2023pal} acquires a simple physical explanation. 
As long as the fundamental space is the momentum space, one can use a dual relation to build spacetime; 
the relative locality is thus a result of parallel transport in momentum space~\cite{Franchino-Vinas:2023}.
Although assigning a central role to momentum space might sound controversial, 
one should remember the customarily construction of quantum field theory:
interactions take place as elastic scattering between particles, i.e. as processes of momentum exchange satisfying energy-momentum conservation.

Third, it is widely-known that one can introduce quantum field theories in noncommutative spaces (see~\cite{Lizzi:2021rlb, Franchino-Vinas:2018jcs, Bevilacqua:2022fbz, DimitrijevicCiric:2023mrn, Hersent:2023lqm} for recent advances). A less studied approach is to introduce the fields and wave equations directly in momentum space\footnote{The older proposals of the Russian school of Gol'fand~\cite{Golfand:1959vqx} and Mir-Kasimov~\cite{Mir-Kasimov:1966a} were actually not fully aware of the connection between their theories and noncommutative spacetimes. }~\cite{Golfand:1959vqx, Mir-Kasimov:1966a, Arzano:2010jw, Franchino-Vinas:2022fkh, Franchino-Vinas:2023rcc}, even though momentum space plays a crucial role in the construction of several theories through Moyal-type products~\cite{Douglas:2001ba}. 
In Refs.~\cite{Franchino-Vinas:2022fkh, Franchino-Vinas:2023rcc} we have brought new elements into the game, such as covariance in momentum space (meaning that observables should be independent of the choice of coordinates in momentum space) and the possibility of introducing Dirac fermions as representations of the Lorentz group in momentum space. In the following we will review our main results in these articles and provide an outlook on how one could develop a proper quantum field theory. 

\section{Klein--Gordon equation in momentum space}
Let us first consider the simpler case of a scalar field $\phi$ of mass $m$. 
In the canonical formalism, its equation of motion is given by the Klein--Gordon equation, which in (flat) Minkowski spacetime with metric $\eta_{\mu\nu}$ reads
\begin{align}
   \left(\eta^{\mu \nu} \frac{\partial}{\partial x^\nu}\frac{\partial}{\partial x^\mu} +m^2\right) \phi(x)\,=\,0\,.
   \label{eq:KG_SR}
\end{align}
Upon Fourier transformation we are lead to an algebraic equation, 
\begin{align}
    (C_{\rm M}(p)-m^2) \tilde \phi (p)\,=\,0\,,
\end{align}
which fixes the on-mass shell in terms of the Casimir
\begin{equation}\label{eq:casimir_M}
  C_{\rm M}(p)\,:=\,p^2\,=\,p_\mu \eta^{\mu\nu}p_\nu\,.
\end{equation}
This Casimir plays at least two roles in Mikowski space. 
On the one hand, together with the square of the Pauli--Lubanski pseudovector, it is a Casimir invariant of the Poincaré algebra. 
On the other hand, since Minskowski space is flat, it also corresponds to the distance from the origin to the point $p^\mu$ in momentum space.
This last interpretation allows for an immediate generalization to rather general, sufficiently smooth, $n$-dimensional curved pseudo-Riemannian manifolds, whose metric will be referred as $g_{\mu\nu}$. 
Taking into account the properties of distance (or Synge's world function~\cite{Synge:1960ueh}), we can write our new ``Casimir'' as
 \begin{align}
 C_\text{D}(p)\,&=\,f^{\mu} g_{\mu \nu }(p) f^{\nu}\,
\label{eq:casimir_metric},
  \end{align}
where the vectors $f^\mu$ play the role of generalized momenta,
\begin{align}
 f^{\mu}  (p)\,:&=\,\frac{1}{2} \frac{\partial C_\text{D}(p)}{\partial p_\mu}\,.
   \label{eq:f_definition}
\end{align} 
The corresponding deformed, algebraic Klein--Gordon equation in curved momentum space then reads 
\begin{equation}
\left(  f^\mu ({p})g_{\mu \nu} ({p})  f^\nu ({p})  -m^2\right) \tilde \phi(p)\,=\,0\,.
   \label{eq:KG_DSR}
\end{equation}
If $\phi$ is realized as a scalar under diffeomorphisms in momentum space, the equation is clearly invariant under redefinitions of the coordinates.

\section{Dirac equation in momentum space}
When trying to generalize the Dirac equation to a curved momentum space we face two problems. The first one is related to the definition of the gamma matrices.
A natural guess is to consider curved-momentum gamma matrices, defined as\footnote{Greek indices denote component of tensors in spacetime/momentum space; Latin ones are used for components in the local orthonormal basis.}
\begin{align}\label{eq:gamma_curved}
 \underline{\gamma}^\mu \,:=\,\gamma^a e^\mu{}_a({p})\,,
\end{align}
written in terms of the usual flat gamma matrices $\gamma^a$ and the vielbein $e^{\mu}{}_a$, 
which in its turn satisfies the relation
\begin{align}
g^{\mu\nu}(p) \,=:\,e^\mu{}_a (p)\eta ^{a b}e^\nu{}_b (p)\,.
\label{eq:metric_tetrad}
\end{align}
As can be explicitly showed from Eq.~\eqref{eq:gamma_curved} and the Clifford algebra satisfied by the flat gamma matrices,
the curved-space gamma matrices satisfy the expected anticommutation relation 
\begin{equation}
\lbrace{ \underline{\gamma}^\mu, \underline{\gamma}^\nu\rbrace}_{}\,=\,2 g^{\mu \nu}(p)\mathbb{1}\,.
\end{equation}

The second concern corresponds to choosing an appropriate momentum. 
In effect, the naive replacement of the momentum with $p_\mu$ yields a noncovariant equation. 
Instead, our proposal is to use the generalized momentum in Eq.~\eqref{eq:f_definition}, 
which by definition is a vector in momentum space, resulting in the following covariant expression 
for the deformed Dirac equation:
  \begin{equation}
 \left( \underline{\gamma}^\mu f_\mu({p})-m\right)\psi(p)\,=\,0\,.
 \label{eq:Dirac_DSR}
\end{equation}
Notice that this is tantamount of introducing a $(1/2, 0)\oplus (0,1/2)$ local representation of the Lorentz group $SO(3,1)$ in momentum space. 
Contrary to what happens in previous attempts\footnote{See Ref.~\cite{Nowicki:1992if} for a fermion in $\kappa$-Poincaré, which was built introducing a finite-dimensional representation in the coproduct of the quantum anti-de Sitter algebra, $SO_q(3,2)$.}, the ``square'' of Eq.~\eqref{eq:Dirac_DSR} equals our deformed Klein--Gordon equation.

\section{Doubly special relativity}
As far as the momentum space is concerned, our discussion until now is valid for arbitrary metrics. 
If instead the underlying momentum space is endowed with a set of symmetries, 
then we can extend our discussions in using these properties.

As a first example, consider a momentum space which admits a set of symmetries whose generators 
leave its origin invariant\footnote{Actually, one could use this invariance to identify an origin.}. 
Then, our generalized Klein--Gordon and Dirac equations will automatically be invariant under this set of symmetries,
thanks to the definition of the Casimir in terms of the distance to the origin.
In particular, this will be the case if the geometry is invariant under the Lorentz transformations.

Another example corresponds to the case in which the $n$-dimensional momentum space possesses a set of $n$ Killing vectors which 
take a given point $q$ into any other point in its neighbourhood. 
We can employ this symmetry of quasi-translations $T$~\cite{Weinberg:1995mt} to define a deformed composition law of coordinates,
\begin{align}\label{eq:modified_composition}
p\oplus q := T_{p}(q),
\end{align}
which is usually understood as a deformed composition of momenta in the noncommutative literature.
Given that this encodes a symmetry of the metric, 
we can write the latter at an arbitrary point $p\oplus q$ in the vicinity of $q$ in terms of the metric at $q$,
\begin{align}
g_{\mu\nu}\left(p \oplus q\right)\,=\, \frac{\partial \left(p \oplus q\right)_\mu}{\partial q_\rho} g_{\rho \sigma }\left( q\right) \frac{\partial  \left(p \oplus q\right)_\nu}{\partial q_\sigma}\,.
\label{eq:iso}
\end{align}
If in addition the metric at the origin $q=0$ reduces to the Minkowski metric and Eq.~\eqref{eq:iso} is globally valid, 
then we can write
\begin{align}
g_{\mu\nu}\left(p\right)\,=\,\left. \frac{\partial \left(p \oplus q\right)_\mu}{\partial q_\rho}\right|_{q\to 0} \eta_{\rho \sigma }  \left. \frac{\partial  \left(p \oplus q\right)_\nu}{\partial q_\sigma}\right|_{q\to 0}\,.
\label{eq:iso_tetrad}
\end{align}
In other words, in this case we obtain a vielbein (or tetrad) which is completely dictated by the deformed composition law~\cite{Carmona:2019fwf}:
\begin{align}
{e}_\mu{}^a (p)\,:=\,\delta^a_\nu \left. \frac{\partial \left(p \oplus q\right)_\mu}{\partial q_\nu}\right|_{q\to 0}\,.
\label{eq:tetrad2}
\end{align}

Our last example is given by maximally symmetric spaces, for which we have at our disposal a set of $n(n-1)/2$ rotational symmetries with generators $J^{\mu\nu}$, as well as $n$ quasitranslations. 
In principle, the quasitranslations are not completely determined, since also the elements
$T^{\prime \rho}:=T^\rho+c^\rho_{\mu\nu}J^{\mu\nu}$, with $c^\rho_{\mu\nu}\in \mathbb{R}$, correspond to isometries in momentum space. 
Taking this into account, the most general sector of translations that the algebra might have (compatible with rotational symmetry) is given by\footnote{Latin indices $i,j$ denotes in this formula spatial indices.}
\begin{align}
[T^0, T^i] \,=\, {c_1}{} T^i + {c_2}{} J^{0i}, \quad\quad\quad [T^i, T^j] \,=\, {c_2}{} J^{ij}\,,
\label{isoRDK}
\end{align}
where $c_{i=1,2}$ are constants. 
As a consequence of this indeterminacy, momentum spaces with the same pseudo-Riemannian metric can be dressed with different composition laws, 
resulting in distinct phenomenological models, as happens for Snyder~\cite{Battisti:2010sr} and $\kappa$-Poincaré~\cite{Kowalski-Glikman:2002oyi},
which correspond both to a de Sitter momentum space.

\section{General discussion of quantum field theory}\label{sec:qft}
One of the main issues of the usual Klein--Gordon equation is the existence of a conserved current whose would-be density has a non-definite sign. 
Indeed, the current $J_{\mu}^{\rm KG}:= {\mathi} \psi^*(x) \overleftrightarrow{\partial}_{\mu} \psi(x)$, where $\overleftrightarrow{\partial}_{\mu}:=\overleftarrow{\partial}_{\mu}-\overrightarrow{\partial}_{\mu}$, defines a sesquilinear form
\begin{align}\label{eq:scalar_KG}
    (\phi,\psi)_{\rm KG}\,:&= \, \mathi \int {\rm d}^3x \big[\partial_t \phi^*(x) \psi(x)-   \phi^*(x) \partial_t\psi(x)\big]\,,
    \\
    &=\,\int \measure \big[ \tilde\phi_+^*(p) \tilde\psi_+(p) -\tilde \phi_-^*(p) \tilde\psi_-(p)\big]\,,
\end{align}
where, after a Fourier transform, we have obtained an expression involving the invariant measure ($\freq:=\sqrt{\vec{p}^2+m^2}$) and have recast it in terms of the positive and negative frequency components of the wave functions:
\begin{align}\label{eq:phi_fourier}
    \phi(x)\,=\, \frac{1}{\sqrt{2}(2\pi)^3}\int \measure \left(e^{\mathi (t \freq + \vec{x}\cdot\vec{p}) }\phi_+(\vec{p}) +e^{-\mathi (t \freq+\vec{x}\cdot\vec{p}) }\phi_-(\vec{p})\right)\,.
\end{align}
It is clear thus that the subset of fields with positive frequency can be used to build the Fock space of the theory~\cite{Wald:1995yp}.

The curved situation is of course more involved~\cite{Wald:1995yp}. Usually one assumes the globally hiperbolicity of the manifold, which is equivalent to saying that the spacetime admits a Cauchy surface.
One could then consider a foliation of the spacetime with Cauchy surfaces parameterized by a normal vector $t_\mu$, 
which can be employed together with the generalized momentum $f^\mu$ to define a (positive) scalar product:
\begin{align}\label{eq:scalar_KG_curved}
    (\phi,\psi)_{ C_{\rm D}}\,:=\, 2\int  {\rm d}^4p \,\sqrt{-g} \, \delta\left(C_{\rm D}-m^2 \right) \,  \Theta(f^{\nu} t_{\nu})\, \phi^*(p) \psi(p) \,.
\end{align}
It satisfies all the necessary conditions to proceed with a Cauchy completion to define a proper Hilbert space and, subsequently, its Fock space. 

We can also introduce actions both for the scalar and fermionic free fields:
\begin{align}\label{eq:KG_action}
S_{\rm KG}\,:&=\,\int {\rm d}^4p\, \sqrt{-g} \,\phi^*(p) \left(  C_{\rm D}(p)  -m^2  \right)  \phi(p)
\,,
  \\
 S_{\rm Dirac}:&=\int  {\rm d}^4p\, \sqrt{-g} \bar\psi(-p)\left(  \underline{\gamma}^\mu f_\mu({p})-m\right)\psi(p)\,.
 \label{eq:Dirac_action}
\end{align}
A variational principle applied to these actions leads to the previous algebraic equations of motion.

One could also try to introduce interactions directly in momentum space, 
introducing terms of higher order in the fields and imposing the conservation of momentum. 
For the latter, there are two natural possibilities: either to choose the conservation of a modified composition law of momenta, cf. Eq.~\eqref{eq:modified_composition},
or of the generalized momenta. Although the latter case would have the advantage of preserving the invariance under diffeomorphisms in momentum space, the distinction between spaces where the acknowledged composition law $\oplus$ is different would be relegated to the vielbein.

\section{Conclusions}
In these lines we have pursued the notion of covariance in momentum space. 
This lead us to consider a deformed Casimir written as a distance in momentum space, cf. Eq.~\eqref{eq:casimir_metric}, 
and define the generalized momenta in Eq.~\eqref{eq:f_definition}. 
This ingredients allows us to introduce the Klein--Gordon and Dirac equations in a natural covariant way, 
using the theory of representations of the Lorentz group in arbitrary momentum spaces. 

If the background momentum space is endowed with some symmetries, these are automatically encompassed by our wave equations. 
The existence of quasitranslations can be used to define a deformed composition law of momentum coordinates,
while in a maximally symmetric momentum space there exists a preferred vielbein.

Next attempts are to be aimed at constructing a consistent quantum field theory. The Lagrangian approach seems rather suitable, taking into account the discussion in Sec. \ref{sec:qft}.
An even more ambitious project would involve considering theories in which a notion of curvature exists also in the spacetime~\cite{Mercati:2023pal, Franchino-Vinas:2021bcl, Franchino-Vinas:2019nqy}.

\ack
The authors are grateful to M.~Arzano, G.~Gori, S.~Mignemi and A.~Trombettoni for helpful discussions. SAF acknowledges the support from Helmholtz-Zentrum Dresden-Rossendorf (HZDR), PIP 11220200101426CO Consejo Nacional de Investigaciones Cient\'ificas y T\'ecnicas (CO\-NI\-CET) and Project 11/X748 (UNLP). 

\bibliography{biblio}

\end{document}